%% file: acl2023.tex
\pdfoutput=1

\documentclass[11pt]{article}

\usepackage[]{ACL2023}
\usepackage{eso-pic}
\usepackage{times}
\usepackage{latexsym}

\usepackage[T1]{fontenc}
\usepackage{amsmath}
\usepackage{mathtools}
\usepackage{xspace}
\usepackage[shortlabels]{enumitem}

\usepackage{breakurl}
\usepackage[breaklinks]{hyperref}
\usepackage{fancyhdr}

\usepackage[utf8]{inputenc}

\usepackage{microtype}
\newcommand{\sepval}{\itemsep-0.5em}
\newcommand{\system}{\ensuremath{\mathsf{SPEC 5G}}\xspace}

\newcommand{\numofsentences}{3,547,587\xspace}
\newcommand{\numofwords}{134M\xspace}

\newcommand{\rawtextamount}{17 GB\xspace}
\newcommand{\simpTestSet}{\ensuremath{\mathsf{5GSum}}\xspace}
\newcommand{\datasetcf}{\ensuremath{\mathsf{5GSC}}\xspace}
\newcommand{\simplocs}{1500\xspace}
\newcommand{\simplocsFinal}{713\xspace}
\newcommand{\numofexperts}{\ensuremath{\mathsf{9}}\xspace}

\newcommand{\numOfSentencesSummarization}{1-12\xspace}
\newcommand{\splitforclass}{85-5-10\xspace}

\newcommand{\numofspecs}{13094\xspace}
\newcommand{\numofwebsites}{13\xspace}
\newcommand{\specialcell}[2][c]{%
  \begin{tabular}[#1]{@{}c@{}}#2\end{tabular}}
\newcommand{\numtotalclass}{2401\xspace}
\newcommand{\numtrainofclass}{2040\xspace}
\newcommand{\numvalofclass}{120\xspace}
\newcommand{\numtestofclass}{241\xspace}

\newcommand{\numclassa}{1303\xspace}

\newcommand{\numclassc}{484\xspace}

\newcommand{\classnamea}{\textit{Non-Security}\xspace}
\newcommand{\classnameb}{\textit{Security}\xspace}
\newcommand{\classnamec}{\textit{Undefined}\xspace}

\newcommand{\bestmodelcf}{{BERT5G}\xspace}

\newcommand{\cfexampleheader}[1]{{\noindent\textbf{Sentence #1:}}\xspace}
\newcommand{\cfexamplelabel}[1]{{\noindent\textbf{Label #1:}}\xspace}

\usepackage{inconsolata}

\title{\system: A Dataset for 5G Cellular Network Protocol Analysis}

\author{
  Imtiaz Karim, Kazi Samin Mubasshir,  Mirza Masfiqur Rahman,  Elisa Bertino \\ Purdue University \\
  \texttt{\{karim7, kmubassh, rahman75, bertino\}@purdue.edu} 
  }

\AddToShipoutPictureBG*{%
  \AtPageUpperLeft{%
    \hspace{\paperwidth}%
    \raisebox{-\baselineskip}{%
      \makebox[0pt][r]{SPEC5G has been accepted for presentation at IJCNLP-AACL 2023 \; \; \; \; \; \; \; \; \;\; \; \; \; \; \; \; 
      \; \; \; \; \; \; \; \;\; \; }
}}}

\begin{document}
\maketitle
\begin{abstract}
5G is the $5^{th}$ generation state-of-the-art cellular network protocol designed to connect virtually everyone and everything with increased speed and reduced latency. Therefore, its development, analysis, and security are critical. 
However, all approaches to the 5G protocol development and security analysis, 
e.g., property extraction, protocol summarization, and semantic analysis of the protocol specifications and implementations are completely manual. 
To reduce such manual efforts, 
in this paper, we curate \system --the \emph{first-ever} public 5G dataset for NLP research. The dataset contains \numofsentences sentences with \numofwords words, from \numofspecs cellular network specifications and \numofwebsites online websites. 
By leveraging large-scale pre-trained language models that have achieved state-of-the-art results on NLP tasks, we use this dataset for security-related text classification and summarization. Security-related text classification can be used to extract relevant security-related properties for protocol testing. On the other hand, summarization can help developers and practitioners understand the high-level idea of the protocol, which is itself a daunting task. To ensure the research community can benefit from this work, all the datasets and accompanying codebase are made publicly available\footnote{Datasets and codebase for \system are publicly available at \href{https://github.com/Imtiazkarimik23/SPEC5G}{https://github.com/Imtiazkarimik23/SPEC5G}}.

\end{abstract}

\maketitle
\input{sections/introduction.tex}
\input{sections/relatedworks.tex}
\input{sections/dataset.tex}

\input{sections/tasks.tex}

\input{sections/experiments.tex}
\input{sections/results.tex}

\input{sections/discussion.tex}

\input{sections/conclusion.tex}
\input{sections/limitation}
\input{sections/ethics.tex}
\section*{Acknowledgement}
\vspace{-0.15cm}
We thank the anonymous reviewers for their insightful comments and the annotators: Adrian Li, Charalampos Katsis, Fabrizio Cicala, Mengdie Huang, Sonam Bhardwaj, Yiwei Zhang, and Zilin Shen from cyber2slab of Purdue University for their valuable time and effort in annotating the downstream task datasets. The work reported in this paper has been supported by NSF under grant 2112471 ``AI Institute for Future Edge Networks and Distributed Intelligence (AI-EDGE)''.



\bibliography{anthology,custom}
\bibliographystyle{acl_natbib}

\appendix


\input{sections/appendix.tex}


\end{document}

%% file: sections/introduction.tex
\section{Introduction}
The deployment of the 5G cellular network protocol has generated a lot of enthusiasm in academia and industry, 
because of its promise of enabling innovative applications, such as autonomous vehicles~\cite{carmap}, remote surgery~\cite{surgery}, industrial IoT~\cite{emergence}, 
augmented reality~\cite{augment}, and multi-player online gaming~\cite{screenbeam}. 
Therefore the security of 5G protocol is critical.
Unfortunately, the 5G protocol development and analysis
are all completely manual tasks requiring 
domain expertise. We observe that for 5G there is an unutilized resource of information available in the form of specifications~\cite{Specifications} 
and numerous tutorials on the Internet. These 
resources have not yet been utilized.

Recently, a few approaches have been proposed that leverage natural language processing (NLP) and machine learning (ML) to detect risky operations in some of the specifications of 4G LTE~\cite{yi_bookwarm} and to analyze change requests~\cite{yi_sr}. These approaches are very limited, not generalizable, and not open-source. Automatic and systematic analysis of 5G networks is still a difficult task. One major problem is the lack of high-quality datasets to train ML models, which would enable the automation of different 5G-related downstream tasks e.g., security-related text classification, protocol summarization, semantic analysis, and automatic programming. 
In this paper, we address this need by introducing \system, a high-quality dataset of the 5G protocol specifications. 5G is not a single wireless technology, but an umbrella term used to categorize the fifth generation of wireless communication, including hundreds of different protocols at different layers of the protocol. Some of these protocols are VoWiFi, 
cellular IoT, IKE, and 5G-AKA. \system is a complete dataset that covers all these protocols and therefore, has the potential to impact different protocols affecting billions of devices. Such a high-quality dataset would be beneficial to numerous applications in different domains, such as security testing, policy enforcement, automatic code generation, and protocol summarization. It would encourage research and development in novel NLP tasks that are communication protocol-specific and critical for the security analysis of these protocols. Notable examples include formal model extraction from large-scale natural language documents and identifications of conflicting security guidelines.

\begin{figure*}[ht]
\centering
\includegraphics[width=0.8\textwidth]{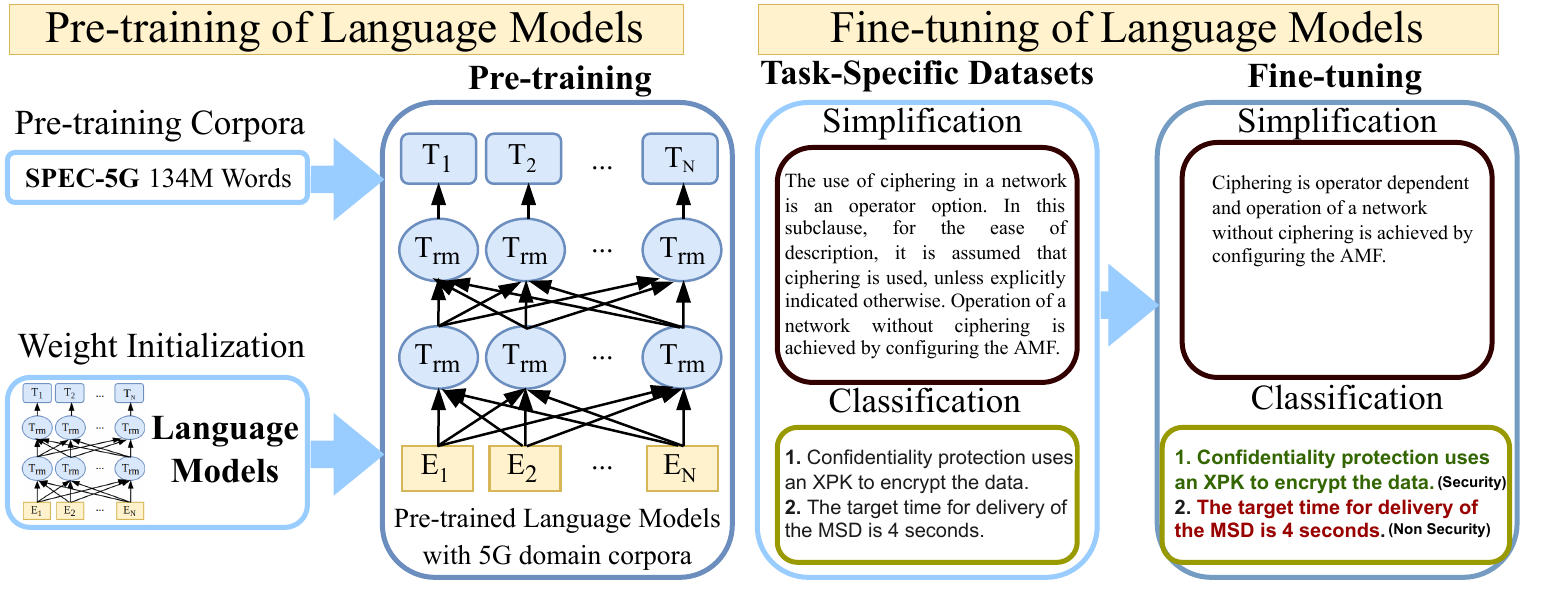}
\caption{Overview of our pre-training and fine-tuning on downstream tasks using \system}
\label{fig:overview}
\end{figure*}
To show the viability of our \system dataset, we use it for two downstream tasks (shown in Figure~\ref{fig:overview}). First, we use it for \emph{security-text classification}. In previous 5G security testing~\cite{basin, cremers2019component, 5greasoner} the properties are manually extracted from the specifications. Using security-text classification, we can automatically identify texts, which specify important security properties to be used for formal verification and other testing approaches. Second, we use \system for the \emph{paragraph summarization} task. The 5G specification is large and complex with specialized jargon, mostly due to backward compatibility requirements. Therefore, it is really daunting for a software developer to understand the high-level ideas of the protocol specification. With the summarization task, we show that it is possible to summarize and simplify the high-level ideas of the protocol. To achieve those tasks, we created two expert-annotated datasets: one for summarization and one for classification. The summarization dataset contains \simplocsFinal long articles and their concise summaries. The classification dataset contains \numtotalclass sentences and their class labels (\classnamea, \classnameb, \classnamec). Both datasets were annotated by multiple
domain experts to ensure quality and fairness. Along with \system, these two expert annotated datasets have been open-sourced to enhance research. 


On the whole, our contributions are three-fold. 
First, we create the first-ever novel 5G dataset (\system) of \numofsentences sentences by pre-processing the 5G specification and scraping data from different 5G tutorials on the Internet. Second, we create two expert-annotated datasets for baseline security-text classification and summarization tasks.
We conduct an extensive evaluation of these datasets using several NLP models on the downstream tasks. The results show that the models pre-trained on \system outperform all baseline models. Third, all these research artifacts have been made available via a public repository. 
To the best of our knowledge, this is the \emph{first-ever} public 5G dataset created for NLP research.




%% file: sections/relatedworks.tex
\section{Related Work}
The introduction of the attention-based transformer architecture by~\cite{DBLP:journals/corr/VaswaniSPUJGKP17} beaconed the era of transformer-based 
Language Models (LM) in the field of NLP. A range of high-performing transformer-based language models have since been proposed, each with its own specific use cases. To train such LMs, high-quality large datasets are critical. 
In the following, we will discuss the research relevant to our work.

\looseness = -1
\noindent \textbf{Cellular Networks Research Using NLP.}
CREEK~\cite{yi_sr} uses BERT models for detecting security-relevant change requests. For this, they pre-train BERT with a subset of 4G LTE specifications 
(1546 out of 13094). Moreover, in ATOMIC~\cite{yi_bookwarm} they design a framework to semantically analyze LTE documents using NLP to obtain a set of hazard indicators for generating test cases based on a given threat model. These are the first steps in applying NLP techniques to analyze cellular network specifications. In a technical blog post from Erricson~\cite{erricson_paper}, the authors adopt LMs for the telecom domain and create a telecom question-answering dataset. Though promising, these approaches do not generalize and are ad-hoc and closed-source,  thus accentuating the need for a complete and public dataset for 5G. 

\noindent \textbf{Summarization.}
Following BookCorpus and Wikidata, researchers have built summarization datasets such as Wikilarge~\cite{zhang-lapata-2017-sentence}, Wikismall~\cite{zhu_wikismall}, and so on~\cite{coster-kauchak-2011-learning, kauchak-2013-improving}. Such datasets are widely used in the field of sentence summarization. Early summarization models mostly relied on statistical machine translation~\cite{wubben-etal-2012-sentence, simplification_narayan}. Improvements of the machine translation model to obtain a new summarization model are done by ~\cite{nisioi-etal-2017-exploring} and investigations on how to simplify sentences to different difficulty levels are conducted after this~\cite{scarton-specia-2018-learning,nishihara-etal-2019-controllable}. Sentence alignment methods to improve sentence summarization are proposed by ~\cite{stajner-etal-2017-sentence} and ~\cite{DBLP:journals/corr/abs-2005-02324}. There are several corpora related to summarization. A large-scale, human-annotated scientific papers corpus is provided by ~\cite{DBLP:journals/corr/abs-1909-01716}. This corpus provides over 1,000 papers in the ACL anthology with their citation networks (e.g., citation sentences, citation counts) and their comprehensive, manual summaries. There is another dataset that has been created for the Computational Linguistics Scientific Document Summarization Shared Task which started in 2014 as a pilot~\cite{jaidka_2014} and which is now a well-developed challenge in its fourth year~\cite{jaidka_2017b, jaidka_2017a}. A new dataset for summarisation of computer science publications by exploiting a large resource of the author-provided summaries is introduced by ~\cite{collins-etal-2017-supervised}.

\noindent \textbf{Sentence Classification.}
The Corpus of Linguistic Acceptability (CoLA)~\cite{warstadt2018neural} consists of English acceptability judgments drawn from books and journal articles on linguistic theory. Each example is a sequence of words annotated with whether it is an English grammatical sentence. The Stanford Sentiment Treebank~\cite{socher-etal-2013-recursive} consists of sentences from movie reviews and human annotations of their sentiment. Sci-Cite~\cite{DBLP:journals/corr/abs-1904-01608} is a large dataset of citation intents for the task of automated analysis of scientific papers by identifying the intent of a citation (e.g., background information, use of methods, comparing results). Researchers have also leveraged other large datasets such as DEFT~\cite{spala-etal-2019-deft} and ACL-ARC~\cite{bird-etal-2008-acl} for the sentence classification tasks. CSABSTRUCT~\cite{cohan-etal-2019-pretrained} is another new dataset of manually annotated sentences from computer science abstracts for Sequential Sentence Classification (SSC). Paper Field~\cite{paper_field} is built from the Microsoft Academic Graph and maps paper titles to one of 7 fields of study: geography, politics, economics, business, sociology, medicine, and psychology. DBpedia is aimed at extracting structured content from Wikipedia. This is a data extract (after preprocessing, with kernel included) with taxonomic, hierarchical categories, or classes, for ~343k Wikipedia articles. A version of this dataset is also a popular baseline for text classification tasks.


%% file: sections/dataset.tex
\section{Dataset Curation}
In this section, we discuss the collection and preparation of our dataset. A significant amount of data was collected from the 3GPP website~\cite{Specifications}.
\subsection{3GPP}
The 3rd Generation Partnership Project (3GPP) is an umbrella organization that hosts several organizations from different countries. 3GPP is globally considered the issuer of standards for cellular network protocols. These standards are publicized as releases, e.g., LTE standards were made public from Release 8 and 5G standards from Release 15. The current release is Release 19.

\begin{figure}[t!]
\includegraphics[width=0.9\linewidth]{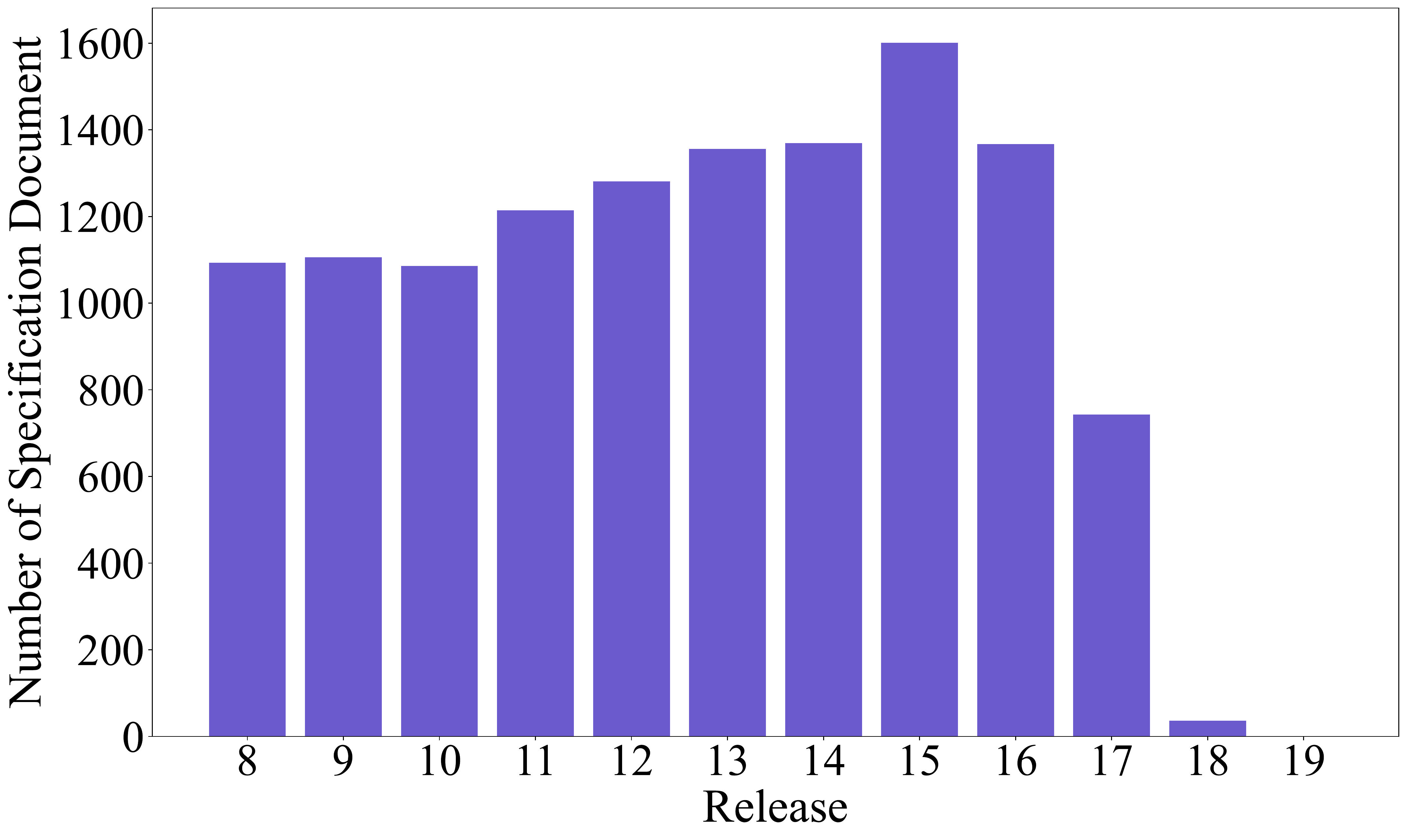}
\caption{Number of Technical Specifications in subsequent releases from 3GPP. Release 15 has seen the largest number of documents. The current release (Rel-19) has only 1 specification document so far. Mean: 1021.08, median: 1160, min: 1, max: 1601, sd: 491.50, skewness: -1.19, kurtosis: 0.13}
\label{fig:SpecVsRel}
\end{figure}

A large number of meeting minutes, Technical Reports (TR) can be found on the 3GPP FTP server ~\cite{Specifications}. 3GPP releases a set of Technical Specifications (TS) as well, which subsequently add features and bug fixes. 
Figure~\ref{fig:SpecVsRel} shows the count of specification documents per release.

\subsection{Dataset Collection}
As stated earlier a significant portion of the dataset has been collected from the 3GPP FTP server.
Automated NLP tasks have been hindered in the 5G domain because of noisy data in the standard documentation. Often, the specification documents contain embedded codes, tables, and lists with definitions of varying terminologies, flow diagrams, finite state machines, and so on--which makes it hard to build models that reason and perform well on downstream applications.

Thus, to leverage downstream NLP tasks, we perform extensive preprocessing. 
Furthermore, we scrape data from \numofwebsites blogs, and forums of the internet. The web sources are listed in Table \ref{Tab:weblist} and details about the web sources can be found in Appendix~\ref{subsec:datasource}.
We extract approximately \rawtextamount of text data from specification releases and web portals using python web scrapper and Selenium~\cite{selenium}. Later we apply a set of standard and domain-specific preprocessing to obtain the final dataset.

\subsubsection{Preprocessing}
5G specifications and web data contain a variety of materials encompassing method and framework documentation, pseudocode, high-level implementations, numerous parameters, field constitution, and so on. At first, the raw data go through standard NLP preprocessing tasks, e.g., removing extra whitespaces, tabs, certain Unicode characters introduced from scrapping, HTML tags, etc.
Later, we extend the preprocessing to handle special cases such as code snippets, tables, figures, references to other specification documents, etc. For the list of preprocessing tasks, we refer the reader to Appendix ~\ref{subsection:preprocessing}. Finally, this dataset is used to pre-train baseline models for downstream applications.

\subsubsection{Dataset Statistics}
\begin{figure}[t!]
\includegraphics[width=0.9\linewidth]{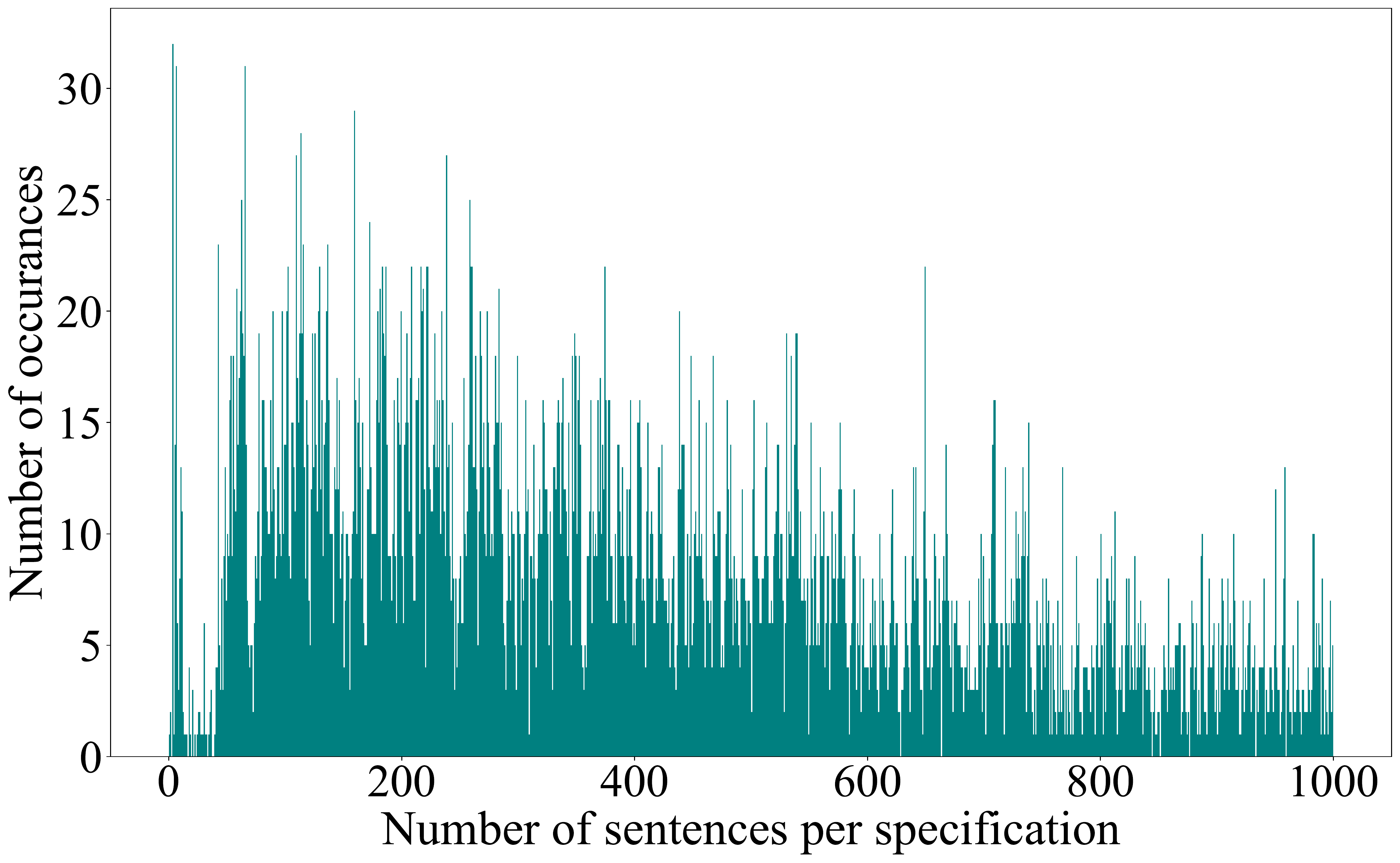}
\caption{Document distribution based on sentences. Documents with more than 1000 sentences were omitted from the figure for better visualization. Mean: 400.90, median: 356, min: 0, max: 1000, sd: 257.83, skewness: 0.52, kurtosis: -0.73}
\label{fig:SentencePerDoc}
\end{figure}

\begin{figure}[ht]
\includegraphics[width=0.9\linewidth]{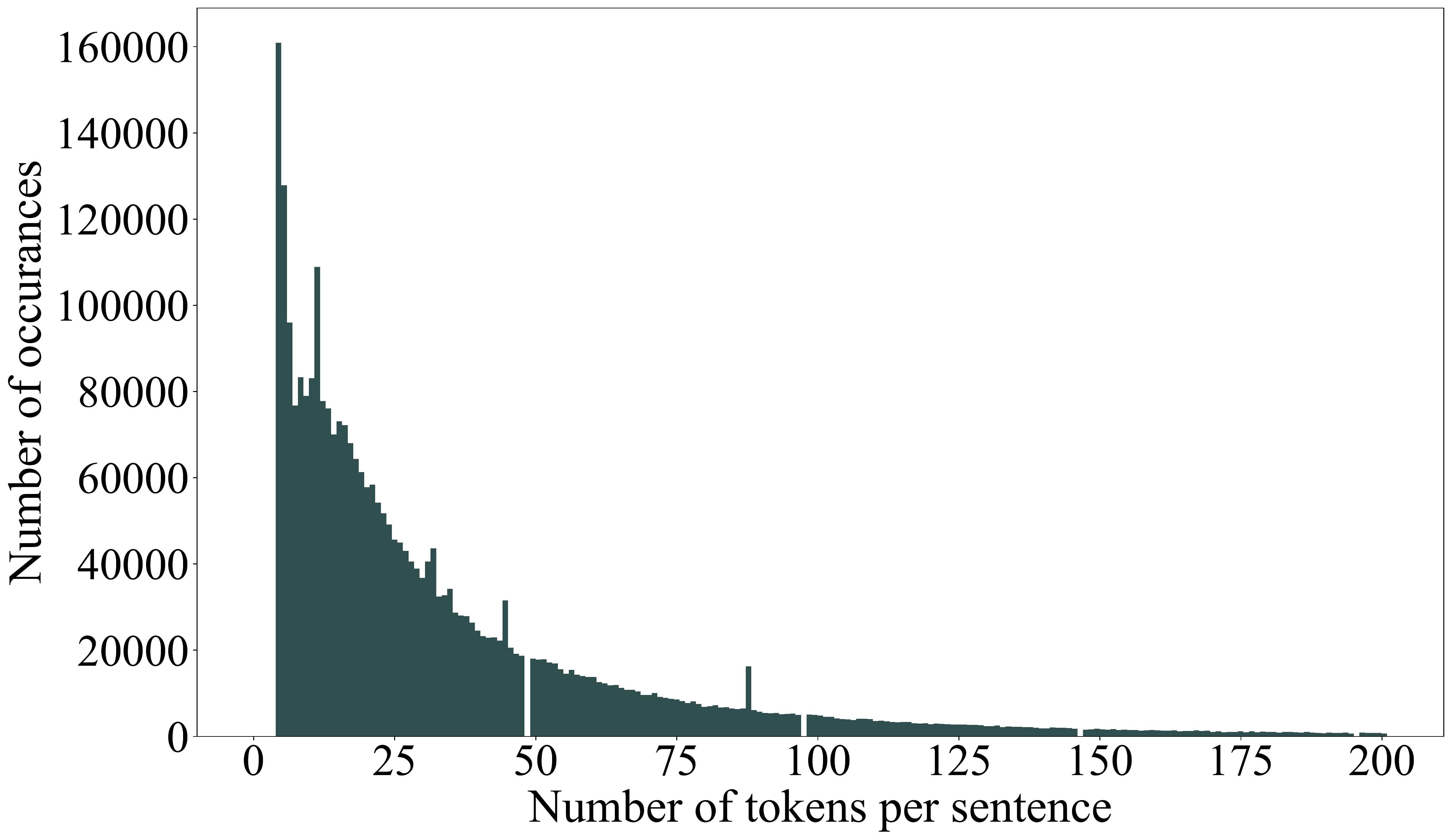}
\caption{Sentence distribution based on tokens. Sentences with more than 200 tokens were omitted from the figure for better visualization. Mean: 34.89, median: 23, min: 4, max: 200, sd: 34.91, skewness: 1.95, kurtosis: 4.10}
\label{fig:TokensPerSentence}
\end{figure}

Our final processed dataset contains \numofsentences sentences with a total of \numofwords words. 
Figure~\ref{fig:SentencePerDoc} shows the distribution of the number of sentences per document and  Figure~\ref{fig:TokensPerSentence} shows the distribution of tokens per sentence.

\begin{figure}[t!]
\includegraphics[width=0.9\linewidth]{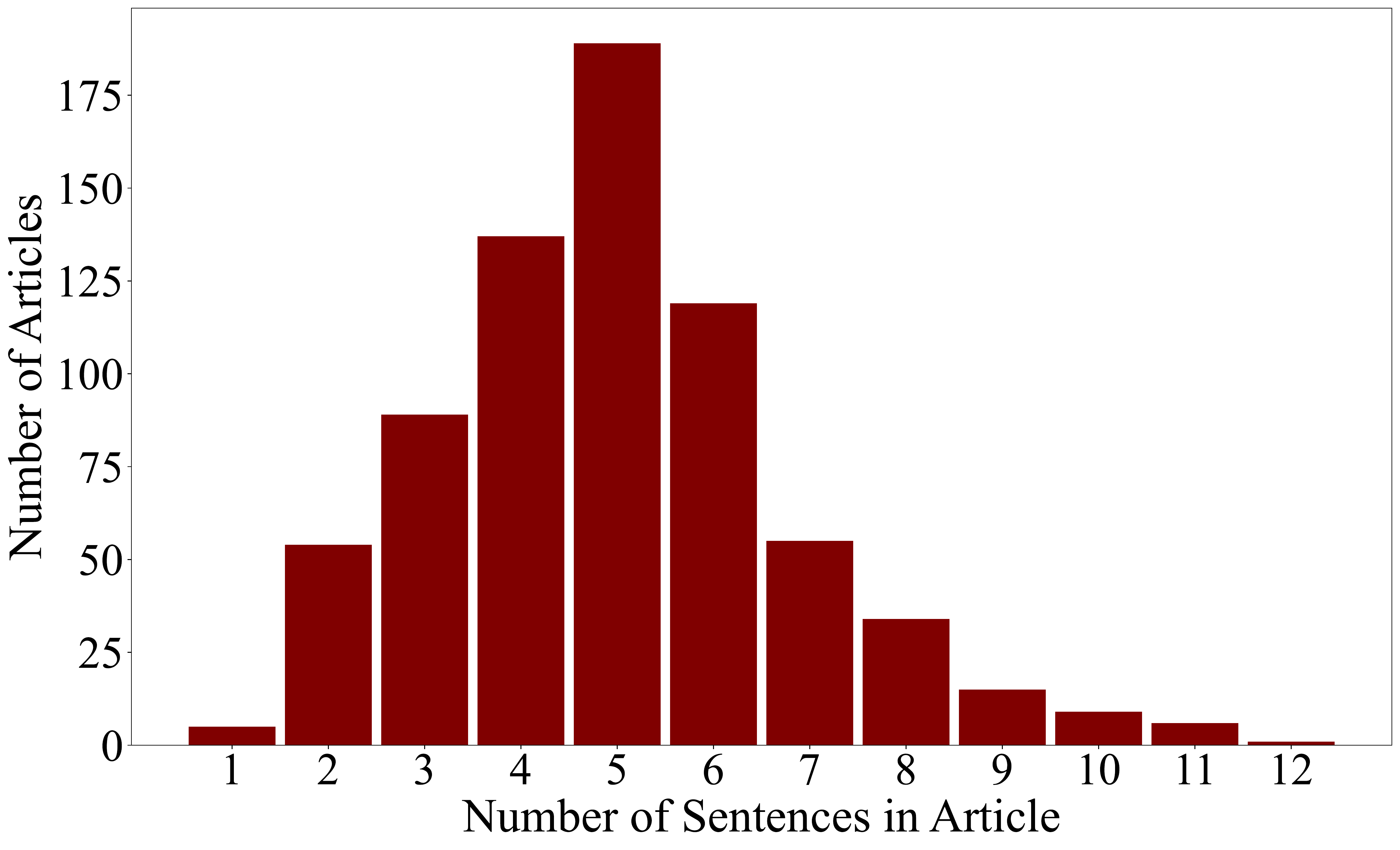}
\caption{Sentence distribution per article. Mean: 4.97, median: 5, min: 1, max: 12, sd: 1.85, skewness: 0.61, kurtosis: 0.72}
\label{fig:SentencePerArticle}
\end{figure}

\subsection{Annotation}
To demonstrate the effectiveness of \system, we additionally create and annotate two datasets specific to two NLP tasks - summarization and sentence classification.

\subsubsection{Summarization}
\label{subsec:sum}

To prepare the summarization dataset, we randomly select \simplocs locations to retrieve articles from the \system dataset. An article is defined as a sequential collection of sentences. Here we apply another round of manual processing to ensure semantic correctness among the sentences of each of the articles. The final curated dataset contains \simplocsFinal articles, each with \numOfSentencesSummarization sentences. The distribution of sentences per article is shown in Figure \ref{fig:SentencePerArticle}. This dataset is subsequently labeled by \numofexperts domain experts; each label itself is a smaller set of sentences that summarizes the article. The task of annotation (summarizing) varies in difficulty. The annotators have made insightful comments about the articles they have faced challenges with. Another round of manual data cleaning has been done based on the comments which resulted in a very high-quality test set for protocol specification summarization. For the rest of this paper, we refer to this annotated dataset as \simpTestSet. 
\subsubsection{Security Classification}
\label{subsec:cf}
Similar to the summarization task, we randomly select and annotate \numtotalclass sentences from our \system dataset to use for multi-class classification. We categorize the data into 3 classes- \classnamea (0), \classnameb (1), and \classnamec (2). To discard human bias, the dataset has been labeled by \numofexperts domain experts. We do a \splitforclass split for train, validation, and test data with \numtrainofclass, \numvalofclass, and \numtestofclass samples respectively. For the rest of the paper, we refer to this dataset as \datasetcf.

Among the 3 classes, the least number of samples are from class 2 (\classnamec: \numclassc).  Yet, the class with the highest number of samples (\classnamea: \numclassa) is about 3 times more than the class with the lowest number of samples. Therefore, the dataset is not highly imbalanced. Overall, this non-uniformity is expected, since most of the specification documents should not be related to \classnameb issues and a high amount of \classnamec statements in 5G specifications would rather mean inconsistencies in implementation.

%% file: sections/tasks.tex
\section{Tasks}
In this section, we define the downstream tasks: summarization and security sentence classification. Moreover, we discuss the relevance of these downstream tasks with respect to 5G.

\subsection{Task 1: Summarization}

Text summarization is the simplification of the original text to a more understandable text while keeping the main meaning of the original text unchanged~\cite{stajner-saggion-2018-data, maddela}. It can provide convenience for non-native speakers~\cite{Petersen2007TextSF, glavas-stajner-2015-simplifying, Paetzold}, non-expert readers~\cite{elhadad-sutaria-2007-mining, siddharthan-katsos-2010-reformulating}. In the case of 5G standard documents, summarization can help developers and practitioners understand the high-level idea of the protocol, which can be really time-consuming without the summarization.

 The document-level text summarization task can be defined as follows. 
Let $C$ be an original complex article; suppose that $C$ consists of $n$ sentences, denoted as $C = {S_1,S_2,\ldots,S_n}$. Document-level summarization aims to simplify $C$ into $m$ sentences, 
which form the simplified article $F$, denoted as $F ={T_1,T_2,\ldots,T_m}$, where $m$ is not necessarily equal to $n$. $F$ retains the primary meaning of $C$ and is more straightforward than $C$, making it easier for people to understand. The operations for sentence-level summarization include word reservation and deletion, synonym replacement~\cite{xu-etal-2016-optimizing}. In our definition, document-level summarization should allow the loss of information but should not allow the loss of important information.  The fact that sentence deletion is a prevalent phenomenon in document summarization is pointed out by \cite{zhong_2020}. We believe that information that has little relevance to the primary meaning should be removed to improve readability.

The objective is to simplify a paragraph without losing important information. Task 1 is more challenging when evaluating a model’s ability to reason about unobserved effects. 

\subsection{Task 2: Sentence Classification}
Text classification is a classic topic for natural language processing, in which one needs to assign predefined categories to free-text documents.
The range of text classification research goes from designing the best features to choosing the best possible machine learning classifiers~\cite{Mekala2021Coarse2FineFT, liu-etal-2021-deep, zhang-etal-2022-learn}. 

The multi-class sentence classification can be defined as follows. Given a sentence $s \in \mathcal{S}$, where $\mathcal{S}$ is some high dimensional sentence space and a finite set of categories or classes $\mathcal{C} = \{c_1, c_2, \ldots, c_n\}$,
the objective of multi-class sentence classification is to find a function $\mathcal{F}$ mapping sentences to categories, formally,
$\mathcal{F}: \mathcal{S} \rightarrow \mathcal{C}$.
Given a dataset $\Tilde{\mathcal{D}}$ of $m$ training samples $\{(s_i, c_i)\}_{i=1}^{m}$, we aim to learn the function $\bar{\mathcal{F}}$ that approximates $\mathcal{F}$.

For protocol analysis, an important step is property-guided testing~\cite{5greasoner}. 
Up to this point, the properties are manually extracted, and the testing is entirely manual. The security classification task aims to label the security-related sentences that in turn can be used as properties and enable semi-automated testing.
 

%% file: sections/experiments.tex
\section{Experiments and Evaluation}

In this section we provide a comprehensive assessment of the proposed methodology through rigorous experimentation and evaluation.
\subsection{Experiment Setup}

\noindent \textbf{Baseline Models:} For baseline models we use base versions of BERT~\cite{devlin2018bert}, RoBERTa~\cite{RoBERTa}, XLNet~\cite{XLNet}, BART~\cite{BART}, GPT2~\cite{GPT2}, T5~\cite{T5}, ALBERT~\cite{ALBERT}, CamemBERT~\cite{CamemBERT}, LongFormer~\cite{Longformer}, Pegasus~\cite{pegasus}; large versions of GPT2 and mBART~\cite{mBART}; medium version of GPT2; DistilGPT2 and DistilBERT~\cite{DistilBERT}.

\noindent \textbf{Pre-trained Models:} 
We pre-train three models- BERT-base, ROBERTa-base, and XLNet-base, on the \system dataset; we refer to them as BERT5G, ROBERTa5G, and XLNet5G respectively. 
The reason for training these three models is discussed in Section \ref{subsection:choiceofmodel} (Choice of Pre-trained Models). 
We then fine-tune the pre-trained models for the downstream tasks. 
The details of the pre-training and fine-tuning are discussed in detail in Section~\ref{training_details}. 

\noindent \textbf{Training Hardware:} We use Google Colab Pro+ to pre-train and fine-tune the models. Around 3000 computing units (CU) of Premium GPU (A100) with high RAM configuration have been consumed to complete all our experiments. For details about CU and training time, we refer to Appendix \ref{sec:CU}.

\subsection{Performance Metric}
To measure the performance of the sentence classification task we use standard performance metrics such as accuracy, precision, recall, and F1-score. We discuss the metrics for the summarization task here in detail.

\subsubsection{Summarization Metrics}
To measure the quality of summarization we use both automatic and human evaluation metrics. For automatic evaluation, we use the commonly used ROUGE score. 

\noindent \textbf{Human Evaluation Metric:} Due to significant dissonance with human evaluation, automatic evaluation metrics are often considered unreliable for summarization quality evaluation. Hence, we resort to human evaluation metrics. The human annotator's rate on a scale from 1 (worst) to 5 (best) on three coarse attributes: (1) \emph{Simplicity:} As the majority of the inferences require speculation, this metric measures how simple and concise the models and the annotators are. (2) \emph{Correctness:} Whether the generated or annotated inferences are grammatically and from a protocol point of view correct. This is very important for the summarization of network protocol specifications. (3) \emph{Contextuality:} Whether the generated or annotated inferences fit the context. 
\begin{table}[t]
\renewcommand{\arraystretch}{1}
\fontsize{8.5}{8.5}\selectfont
\begin{tabular}{p{0.3\linewidth}p{0.15\linewidth}p{0.15\linewidth}p{0.15\linewidth}} 
 \hline
 Model & RG-1 & RG-2 & RG-L \\ [0.5ex] 
 \hline\hline
 BERT-base & 0.484 & 0.341 & 0.415 \\
 \hline
 BERT5G & \textbf{0.543} & \textbf{0.382} & \textbf{0.472} \\
 \hline
 PEGASUS & 0.239 & 0.120 & 0.199 \\
 \hline
 RoBERTa-base & 0.489 & 0.341 & 0.418 \\
 \hline
 RoBERTa5G & 0.540 & 0.379 & 0.469 \\
 \hline
 BART-base & 0.357 & 0.231 & 0.311 \\
 \hline
 XLNET-base & 0.483 & 0.340 & 0.416 \\
 \hline
 XLNET5G & 0.526  & 0.362 & 0.453 \\
 \hline
 GPT2 & 0.488 & 0.340 & 0.418 \\
 \hline
 GPT2-medium & 0.481 & 0.333 & 0.408 \\
 \hline
 GPT2-large & 0.487 & 0.344 & 0.418 \\
 \hline
 DistilGPT2 & 0.483 & 0.333 & 0.412 \\
 \hline
 T5 & 0.444 & 0.285 & 0.363 \\
 \hline
\end{tabular}
\caption{Performance of summarization over the \simpTestSet dataset based on ROUGE scores.}
\label{tab:summarization_performance}
\end{table}


\subsection{Performance}
\label{sec:performance}
We now report the performance of the baseline language models to characterize our dataset. 

\noindent\textbf{Summarization:}
We report the models' mid-score fmeasure of ROUGE-1, ROUGE-2, and ROUGE-L in Table~\ref{tab:summarization_performance} and human evaluation scores in Table~\ref{tab:human_eval} to show the comparison of the baseline models with the pre-trained models. BERT5G outperforms all the models, though BERT-base was not the best-performing model. This shows the quality of our dataset for technical specification learning.

\noindent\textbf{Sentence Classification:}
We report the performance of the models on the sentence classification task in Table~\ref{tab:Finetuning_performance}. For sentence classification with relatively few classes, BERT, ROBERTa, and XLNet perform the best\cite{Chang2020}. Therefore, we pre-train 3 models- BERT-base, ROBERTa-base, and XLNet-base language models on the \system dataset. These 3 models along with other baselines are then fine-tuned on the \datasetcf dataset to compare their classification performance. We observe that BERT5G, ROBERTa5G, and XLNet5G outperform their corresponding baselines by a significant margin. Additionally, BERT5G outperforms all other models in precision and F1 score. XLNet5G has the highest recall. Interestingly, the baseline GPT2 has the highest accuracy. Despite that, we do not choose GPT2 for pre-training. The reason behind this is discussed in Section~\ref{result analysis}.



\begin{table}[ht]
\renewcommand{\arraystretch}{1}
\fontsize{9}{9}\selectfont
\begin{tabular}{p{2cm}cccc} 
 \hline
 Model & Precision & Recall & F1 & Acc\\ [0.5ex] 
 \hline\hline
 ALBERT-base & 0.6485 & 0.6587 & 0.6459 & 0.7034 \\
 \hline
 BART-base & 0.6458 & 0.6503 & 0.6432 & 0.7103 \\
 \hline
 BERT-base & 0.6113 &	0.6229 & 0.6157	& 0.6897 \\
 \hline
 BERT5G & \textbf{0.6972} & 0.6762 &  \textbf{0.6856} & 0.7655 \\
 \hline
 CamemBERT & 0.6107 & 0.6272 & 0.6174 & 0.7034 \\
 \hline
 DistilBERT & 0.5819 & 0.5769 & 0.5731 & 0.6621 \\
 \hline
 GPT2 & 0.6133 & 0.5567 & 0.5767 & \textbf{0.7973} \\
 \hline
 LongFormer & 06280 & 0.6281 & 0.6274 & 0.7034 \\
 \hline
 mBART-large & 0.6598 & 0.6642 & 0.6606 & 0.7241 \\
 \hline
 ROBERTa-base & 0.5752 & 0.5562 & 0.5631 & 0.6690 \\
 \hline
 ROBERTa5G & 0.5944 & 0.5696 & 0.5785 & 0.6966 \\
 \hline
 XLNet-base & 0.6260 & 0.6339 & 0.6297 & 0.7034 \\
 \hline
 XLNet5G & 0.6480 & \textbf{0.6829} & 0.6619 & 0.7103 \\
 \hline

\end{tabular}
\caption{Performance of baseline models on classification task over the \datasetcf dataset.}
\label{tab:Finetuning_performance}
\end{table}

%% file: sections/results.tex
\begin{figure*}
\fbox{
\renewcommand{\arraystretch}{1}
\fontsize{8}{8}\selectfont
\parbox{\textwidth}{
\textbf{Article:} The PDCCH channel is used to send control information to the UEs on a per-slot basis. It is used to schedule downlink transmissions, uplink transmissions, MCS of those transmissions, and HARQ information. The PDCCH can appear on any subcarrier; so the jammer must decode the parameter CORESET freq-domain. The parameter CORESET-time-duration, which can take on values 1, 2, or 3, indicates how many OFDM symbols the PDCCH occupies in each slot. The PDCCH always starts in the first symbol of each slot, is QPSK modulated and uses polar coding.
\vspace{0.01\linewidth}

\textbf{Reference:} PDCCH sends per-slot control information to UEs. Schedules downlink, uplink, MCS, and HARQ transmissions. PDCCH can exist on any subcarrier; the jammer must decode CORESET freq-domain. The CORESET-time-duration parameter shows how many OFDM symbols the PDCCH occupies each slot. The PDCCH starts at the first symbol of each slot and uses polar coding.
\vspace{0.01\linewidth}

\textbf{BERT-base:} The PDCCH channel is used to send control information to the UEs on a per-slot basis. \textcolor{brown}{\textbf{The PDCCH can appear on any subcarrier; so the jammer must decode the parameter CORESET freq-domain.}}
\vspace{0.01\linewidth}

\textbf{BERT5G:} The PDCCH channel is used to send control information to the UEs on a per-slot basis. \textcolor{teal}{\textbf{The parameter CORESET-time-duration, which can take on values 1, 2, or 3, indicates how many OFDM symbols the PDCCH occupies each slot.}}
\vspace{0.01\linewidth}

\textbf{RoBERTa-base:} The PDCCH channel is used to send control information to the UEs on a per-slot basis. \textcolor{brown}{\textbf{The PDCCH can appear on any subcarrier; so the jammer must decode the parameter CORESET freq-domain.}}
\vspace{0.01\linewidth}

\textbf{RoBERTa5G:} The PDCCH channel is used to send control information to the UEs on a per-slot basis. \textcolor{teal}{\textbf{The parameter CORESET-time-duration, which can take on values 1, 2, or 3, indicates how many OFDM symbols the PDCCH occupies each slot.}}
\vspace{0.01\linewidth}

\textbf{XLNet-base:} The PDCCH channel is used to send control information to the UEs on a per-slot basis. The PDCCH always starts in the first symbol of each slot, is QPSK modulated and uses polar coding.
\vspace{0.01\linewidth}

\textbf{XLNet5G:} The PDCCH channel is used to send control information to the UEs on a per-slot basis. The PDCCH always starts in the first symbol of each slot, is QPSK modulated and uses polar coding.
}
}
\vspace{-0.2cm}
\caption{Comparison of summarization task by pre-trained models and their base version. \textcolor{brown}{\textbf{Brown}} colored lines denotes the base models inability to capture protocol specific sentences in summaries and \textcolor{teal}{\textbf{teal}} colored lines donotes the sentences introduced by pre-trained models on \system that are more contextual.}
\label{fig:human_evaluation}
\end{figure*}

\section{Result Analysis}\label{result analysis}

\noindent \textbf{Performance Improvements Due to \system:} The primary objective of our work is to introduce an anchor 5G dataset that might pave the way for future NLP research in 5G and NLP. The models pre-trained on 5G and fine-tuned on respective tasks achieve significant performance improvements, suggesting that such dataset should be considered the gold standard for pre-training models before deploying them for more sophisticated, 5G-oriented NLP applications.

\noindent \textbf{Best Pre-trained Model:} The scores of both tasks show that \bestmodelcf is the best-performing model. It is not surprising that XLNet5G, the pre-trained version of more recent BERT variant XLNet, is a close competitor. While GPT2 is a good choice for our summarization task, we do not recommend it for security classification. Despite a high accuracy score, GPT2 failed to achieve a contending recall or F1-score.
We observe that GPT2 could classify the \classnamea samples well (53 out of 70 test samples were correct) which dominates the dataset distribution, while poorly classifying samples from \classnameb (11 out of 24 correct) and \classnamec (4 out of 11 correct). This is the reason for its higher accuracy yet low recall and F1. 

\noindent \textbf{Results of Human Evaluation for Summarization:} We randomly sample 40 inferences generated by each pre-trained model, their non-pre-trained versions, and corresponding gold inference. These inferences are then manually rated by three independent annotators based on the human-evaluation metrics. As shown in Table \ref{tab:human_eval}, we observe that the fine-tuned models perform similarly on \system but fail to reach gold annotation performance. Moreover, as expected, the pre-trained models significantly outperform their non-pre-trained counterparts. We provide some examples of the generated inferences in Figure~\ref{fig:human_evaluation}. 
Inspection of the model-generated inferences reveals that the usage of keywords from the technical specifications is more frequent in inferences generated by models pre-trained on \system.

\begin{table}[ht]

\renewcommand{\arraystretch}{1}
\fontsize{9}{9}\selectfont
\begin{tabular}{cccc}
 \hline
Model & Simplicity & Correctness & Contextuality \\
\hline 
\hline 
Gold & 4.37 & 4.77 & 4.56 \\
\hline 
BERT-base & 3.2 & 3.87 & 3.3  \\
\hline
BERT5G & 3.96 & 4.32 & 3.92 \\
\hline 
RoBERTa-base & 3.7 & 4.03 & 3.76 \\
\hline
RoBERTa5G & 4.02 & 4.2 & 3.94 \\
\hline 
XLNET-base & 3.57 & 3.84 & 3.53 \\
\hline 
XLNET5G & 3.97 & 4.31 & 3.91 \\

\hline

\end{tabular}
\caption{Result of the human evaluation for \system.}
\label{tab:human_eval}
\end{table}

%% file: sections/discussion.tex
\section{Discussion}

\noindent \textbf{Broader Impact of \system.}
Our dataset can offer valuable insights and applications that extend beyond the immediate scope. It can be very useful also for other specialized communication protocols (like IoT, Bluetooth, Bluetooth Low Energy, Vehicular Protocols, and WiFi). One popular methodology to evaluate the design of communication protocols is to manually extract a formal model, for example in terms of finite state machines, of the protocol and evaluate the model against the desired security and privacy properties~\cite{5greasoner}.
One major issue with this approach is that the manual model extraction from the protocol-text 
is error-prone and not scalable. Therefore, communication protocols are analyzed partially or within a specific scope. The analysis paved by \system can yield a deeper understanding of network behavior, interference patterns, and potential optimizations that can be applied to a variety of wireless communication scenarios. This can lead to an ecosystem around our initial dataset, which could include models trained on \system and additional relevant datasets that could be combined with ours.

The versatility can spark innovation in the design and development of future-generation protocols. There is a lot of work to design technologies and protocols for interconnecting cellular networks and non-terrestrial networks (via for example, low earth orbit--LEO satellites); therefore creating datasets for these new protocols would only require addition to a modest amount of new data to \system. Similarly, models trained on \system could also be tuned by using a modest amount of new data. By leveraging the insights gained from the interactions within the dataset, researchers and engineers can create protocols that can adapt to evolving communication landscapes. This adaptability will be essential as we move towards more interconnected and heterogeneous networks.

Furthermore, the utility of our dataset reaches beyond those exclusively working on 5G networks. As NLP research and natural language processing techniques continue to evolve, our dataset can serve as a foundation for various research avenues. For instance, the dataset can be employed to develop advanced predictive models, anomaly detection systems, and intelligent network management solutions. These applications are not limited to the realm of 5G but have the potential to influence and enhance NLP research across a broader spectrum.

\label{subsection:choiceofmodel}
\looseness = -1
\noindent \textbf{Choice of Pre-trained Models.} 
The motivation behind the choice of the machine learning models is to show the quality of \system. Hence we only use pre-existing models for the downstream tasks and do not measure the performance of simple baseline models like lead-3 extractive baseline (taking the first 3 sentences of the article as the summary) and the SummaRuNNer extractive model~\cite{DBLP:journals/corr/NallapatiZZ16}, nor improve the performance of the downstream tasks. We pick the models that perform well in both downstream tasks. Here the chosen models are all encoder-only to maintain consistency between the experiments. Nevertheless, encoder-decoder models or decoder-only models would also benefit from pre-training.
It is well known that pre-training on domain-specific data can help to improve the performance of downstream tasks in the domain~\cite{biobert, legalbert}.
However, in our case after the first step of preprocessing, BERT only improves 2.73\%, XLNet improvement is 1.96\% and ROBERTa improves 0.061\% in F1 score. Thus, although we commonly know that pre-training improves downstream tasks, evidently the preprocessing of the dataset signifies that process even more. The performance improvement of the base models after pre-training on our dataset indicates that the models could learn and sufficiently generalize their knowledge in technical specifications. 
We leave exploring the downstream tasks in detail and the criteria for the selection of different models on the technical specification domain as future work.

\noindent \textbf{Downstream Task Dataset Size.}
While the downstream task datasets may seem small, recent high-quality manually annotated datasets 
had similar sizes--COUGH dataset(1236 labeled sentences)\cite{zhang-etal-2021-cough} and YASO dataset (2215 labeled sentences)\cite{orbach-etal-2021-yaso}. Thus, the current size is comparable to the contemporaries. To address the selection bias of the relatively short test set, the test points are randomly sampled on 3 different runs of each model, and the models are run on 3 different random seeds which show low standard deviation in performance metrics. Therefore, the randomness in the test set removes the selection bias. Moreover, this dataset can easily be used as a seed alongside our trained models for semi-automatic annotation with minimal human effort. Our work enables this direction of using language models in technical specification documents. In the case of the summarization dataset, it is only used as a test set for the models that can already summarize articles. Their performance on summarizing network protocol specification is measured using this test set. 

\noindent \textbf{Project Maintenance.}
In the context of the 3GPP, major releases like 3G, 4G, and 5G are published every ten years. However, smaller, incremental functional changes are made each year within these larger frameworks. These updates are designed to be backward compatible and avoid conflicts with the previous releases, ensuring a smooth transition for existing infrastructure and devices. To address the concern with the new releases, we have devised a plan to maintain the quality and relevance of our dataset in the face of these protocol changes. After each major 3GPP release, we will analyze the changes and updates made to the protocol specifications. For each significant protocol update, we will review and re-evaluate the annotations in our dataset. This will involve identifying any modifications, additions, or clarifications in the protocol specifications. The Change Request (CR) procedure used by 3GPP to create revised versions of 3GPP specifications can be used to automatically identify the modifications. Our team will then update the dataset to accurately reflect these changes. Alongside these updates, we will provide summaries of the changes made in each protocol release. This will serve as a "TL;DR" version highlighting the key modifications that have taken place. This way, users can quickly understand what has changed in the context of each new release. Our commitment is to keep our dataset in sync with the evolving protocols and maintain its utility as a valuable resource for researchers, developers, and industry professionals.

%% file: sections/conclusion.tex
\section{Conclusion}
We have created \system --a new dataset for 5G, \simpTestSet and \datasetcf -- expert annotated datasets for 5G protocol summarization and 5G security text classification respectively. To show the usefulness of \system in protocol specifications learning by the Language Models, we design security sentence classification and summarization tasks for state-of-the-art Language Models to solve.

\noindent \textbf{Future Work.}
Given the specialized nature of 5G terminology in the dataset, it could be utilized for domain adaptation tasks in NLP (for instance, adapting language models to understand and generate content in the context of 5G communications). The dataset could be used to create datasets for named entity recognition tasks, focusing on extracting and categorizing specific entities such as protocols, technologies, companies, and standards relevant to 5G. With the wealth of information present in the dataset, question-answering datasets could also be constructed, where models are trained to answer questions related to 5G concepts, protocols, and technologies. \system can be used to develop semantic role labeling datasets, assisting in understanding the roles and relationships of various elements in sentences discussing 5G. Datasets for document classification tasks, where the goal is to categorize entire documents or articles based on their content related to 5G concepts can also be created. With content from various sources, the dataset could be used to create parallel corpora for translating technical 5G content between different languages. \system can be utilized to develop datasets for dependency parsing tasks, improving syntactic analysis and understanding of relationships between words. Generating datasets for topic modeling tasks can help in identifying and categorizing prevalent topics within the 5G domain.


%% file: sections/limitation.tex
\section{Limitations}


Here we discuss some limitations we faced.
\subsection{Underspecifications in the standards}
In this paper, we introduce \system, a dataset aimed at the automated analysis of the 5G protocol, and show the usefulness of \system in two downstream tasks. The performance of the two different downstream tasks on the dataset, in turn, depends on the 5G standards. In some cases, the standards are intentionally kept underspecified and contain ambiguities. The reason for such underspecifications and ambiguities is mainly to give vendors flexibility in the implementation design and performance enhancement. Nonetheless, the \system dataset can include some of the underspecified behaviors from the standards. These ambiguities existing in the text can be resolved using human expertise. This is precisely how we leverage human expertise for the two downstream tasks in the paper. However, this can be accomplished by using NLP methods that exploit unlabeled data and human knowledge. This is the direction we plan to pursue in the future.

\subsection{Automation} The aim of \system is to help automate the manual-intensive tasks of 5G protocol development, analysis, and testing using state-of-the-art NLP techniques. However, it is evident that it is still not possible to completely automate such tasks because of the manual annotation, which requires domain expertise. In spite of the limited annotated data, we show that it is still possible to achieve fairly good results in two downstream tasks. It may not be possible to completely automate the 5G related tasks,  but we still hope it can help reduce the large manual efforts which is the current state-of-the-art.

%% file: sections/ethics.tex
\section{Ethical Considerations}
\vspace{-0.15cm}
In regards to the datasets being released, all information is in the public domain and is not subject to any copyrights. 
To pre-train, we use different language models. It has been reported that the pre-trained masked language models encode unfair social biases such as gender, racial bias, and religious biases~\cite{bommasani2020interpreting}. In our case, as we are dealing with a technical domain, we believe these biases do not have any impact on our results. Moreover, we randomly evaluated the model's outputs and found no evidence of these biases.
In the case of annotations, the annotators for \system are all Ph.D. students doing active research in the area of networks. They are provided with specific guidelines (discussed in detail in Appendix~\ref{guidelines}) and are strictly asked not to write any toxic content (hateful or offensive toward any gender, race, sex, or religion) and to consider gender-neutral settings.

%% file: sections/appendix.tex
\noindent \textbf{Appendix}
\section{Preprocessing and Training}
In this section, we outline the preprocessing steps undertaken to clean and transform the data, as well as the training methodology employed to optimize the model's performance. By employing rigorous preprocessing and training techniques, we aim to ensure reliable and accurate results in our subsequent analysis.
\subsection{Preprocessing Details}\label{subsection:preprocessing}
 The following preprocessing steps were performed-
\begin{itemize}
\sepval
    \item Sentences containing codes (e.g. consecutive '\{\{', '\}\}', '((', '))') are removed.
    \item Some of the remaining html tags present after web scrapping are removed.
    \item Citations and references are removed.
    \item Sentences mentioning subsequent figures, and tables are removed.
    \item Figure/table captions are skipped.
    \item Sentences containing consequent digits and dots refer to (sub)section headers are removed.
    \item Some malformed texts appearing from figures and tables after parsing text files from .doc or .pdfs files are filtered out.
    \item Sentences containing Unicode characters appearing as raw texts are removed.
    \item Multiple consecutive newlines, tabs, whitespace, and delimiters are processed into one.
    \item Starting numbers, dots, interpuncts, and hyphens appearing from (un)ordered lists are removed.
    \item Additional whitespaces after opening parentheses, curly braces, and brackets are removed. Similar to closing ones.
    \item 3GPP specifications contain numerous mentions of specification documents (i.e. TS 24.301). These do not add any useful features for learning. Those are renamed as "specification document".
    \item If a sentence contains a high amount of digits, they necessarily are from embedded codes. If more than 20\% are seen, we filter out the sentence.
    \item Few special cases (for example: "e.g.,", "i.e.,") are handled to not be considered as the end of a sentence.
    \item An additional newline is added after adding all texts from each of the documents/web pages. This is to ensure that certain downstream applications (e.g., summarization) do not get affected by unrelated texts from multiple documents. 
\end{itemize}

\subsection{Training Details}
\label{training_details}
To pre-train BERT Masked Language Model (MLM), we use the Adam optimizer with $\epsilon = 10^{-8}$ and train the model for $10$ epochs. The learning rate is $5\times 10^{-5}$, we set aside $10\%$ of the data as validation to inspect the model performance at every 50k steps. BERTFastTokenizer has been used to tokenize the dataset. We use the same parameters to pre-train ROBERTa MLM for $5$ epochs and ROBERTa BPE tokenizer to tokenize the dataset for this setting. For pre-training XLNet Permutation Language Model (PLM), we use the Adam optimizer in the same setting. Since XLNet requires approximately $5$ times more than BERT or ROBERTa, we train the model for $1$ epoch. We use the SentencePiece tokenizer in this case.

When fine-tuning the classification models, we set the learning rate to be $2\times 10^{-5}$, weight decay to be $0.01$, and batch size to be $16$. The Huggingface standard pipeline with the Automodel class has been used for sequence classification. We train each model for $15$ epochs.

We use the bert-extractive-summarizer~\cite{bert-extractive-summarizer} to generate summaries using BERT-base. The Huggingface standard pipeline libraries has been used to generate summaries using sequence-to-sequence models i.e., PEGASUS and T5 that comes with default summarization capability. To generate summary using RoBERTa-base, RoBERTa5G, XLNet5G, and BERT5G, we use the Huggingface Automodel Library. We use another Huggingface library TransformerSummarizer to generate summary using XLNet, GPT2, GPT2-base, GPT2-medium, GPT2-large and DistilGPT2.

\subsection{Compute Unit and Training Duration}
\label{sec:CU}
 A compute unit (CU) is the unit of measurement for the resources consumed. To calculate CUs, one needs to multiply two factors: (1) Memory (GB) - size of the allocated server for task to run and (2) Duration (hours) - how long the server is used. This means, 1 CU = 1 GB memory x 1 hour. We have used around 80-90\% of the GPU during training time. By definition of computing units, we have used roughly 100 hours of 30GB GPU time.

Pre-training BERT takes around 36 hours in our experimental setup. Pre-training RoBERTa and XLNET takes around 24 hours each. Fine-tuning each model takes around 5-6 hours. 

\section{Performance Evaluation of Downstream Tasks}
We report the evaluation of performance and metrics used for it in this section.
\subsection{Performance Metrics}
For automatic evaluation, we use the commonly used ROUGE score. We use the Python rouge\_score library to calculate this. 
\noindent \textbf{ROUGE Score:}\label{sec:rog_score}
ROUGE-N measures the number of matching ‘n-grams’ between the model-generated text and a ‘reference’. An n-gram is simply a grouping of tokens/words. A unigram (1-gram) would consist of a single word. A bigram (2-gram) consists of two consecutive words. In ROUGE-N, N denotes the n-gram that is being used. For ROUGE-1 the match rate of unigrams between the model output and reference are measured. ROUGE-2 and ROUGE-3 would use bigrams and trigrams respectively.

\noindent \textbf{Recall:} The recall counts the number of overlapping n-grams found in both the model output and reference, then divides this number by the total number of n-grams in the reference. 

\begin{equation*}
    recall = \frac{count_n(gram_n)}{count(gram_n)}
\end{equation*}

\noindent\textbf{Precision:}
We use the precision metric — which is calculated in almost the exact same way as recall, but rather than dividing by the reference n-gram count, it is divided by the model n-gram count.

\begin{equation*}
    precision = \frac{num\ of\ ngrams\ in\ model\ \&\ ref}{num\ of\ ngrams\ in\ model}
\end{equation*}

\noindent Now that both the recall and precision values are available, they can be used to calculate the ROUGE F1 score with the following formula:

\begin{equation*}
    2 * \frac{precision * recall}{precision + recall}
\end{equation*}

\noindent\textbf{ROUGE-L:}
ROUGE-L measures the longest common subsequence (LCS) between the model output and the reference. With this metric, the number of tokens in the longest sequence shared between both are counted. The idea here is that a longer shared sequence would indicate more similarity between the two sequences. The recall and precision calculations can be applied just like before — but this time the match is replaced with LCS.

\begin{equation*}
    recall = \frac{LCS(gram_n)}{count(gram_n)}
\end{equation*}


\section{Annotation Guidelines}
\label{guidelines}
Below are the specific guidelines that we have given to the annotators to ensure the standard of annotation.
\subsection{Sentence Classification Guidelines}
The annotators are given some general guidelines and are suggested to follow some steps to make the data annotation consistent. They are also provided with some rules and tips.\\
\noindent\textbf{General Guidelines: }
For this task, an annotator is given a set of sentences. Based on the methods, fields, variables, and/or entities mentioned in the sentence, the annotator's objective is to identify if the sentence implies a potential security concern or sophisticated operation that might involve vulnerable consequences. \\
\noindent\textbf{Steps: }
\begin{enumerate}
\sepval
    \item Read the sentence carefully.
    \item Identify items and the operation that involve a security issue.
    \item Decide the label based on the following:
    \begin{enumerate}[a)]
        \item \textcolor{teal}{\classnamea}: The expressed operation cannot be exploited/ The sentence does not describe any security hazard/ does not describe any underspecified criteria/ does not involve complicated, flawed operations.
        \item \textcolor{red}{\classnameb}: The text discusses situations or operations that might be risky/ The text involves certain properties or variables, exploiting which, one can seriously block the operations, or harm the entity, or breach privacy.
        \item \textcolor{blue}{\classnamec}: The discussed operation is not clear/ The sentence does not express all the parties or variables involved/ The sentence entails some previous operation unavailable to the annotator- without which the annotator can not decide about the potential risk.
    \end{enumerate}    
    \item If the sentences do not express any proper context or are semantically incorrect, or have no items with a sentiment expressed towards them, add a comment and proceed to the next data.

\end{enumerate}

\textbf{Rules and Tips: }
\begin{itemize}
\sepval
    \item Select all items in the sentence that have a
    security hazard.
    \item If there are multiple such cases, you may choose any or all of them.
    \item Optionally, you may provide a comment about your rationalization or feedback about the data (e.g., errors, unclear descriptions.)
\end{itemize}

\subsection{Summarization Guidelines}
Similar to classification guidelines, the annotators are given general guidelines and suggested steps to annotate the summarization dataset. Again, they are also provided with some rules and tips. Below are the guidelines for annotation tasks for summarization. \\ 
\noindent\textbf{General Guidelines: }
For this task, an annotator is given a set of articles. Based on the methods, fields, variables, and/or entities mentioned in the sentence, the annotator's objective is to summarize the article without losing important information, correctness and contextuality. \\
\noindent\textbf{Steps: }
\begin{enumerate}
\sepval
    \item Read the article carefully.
    \item Identify the key points.
    \item Summarize the article by doing the following:
    \begin{enumerate}[a)]
        \item \textcolor{teal}{Deletion}: Delete a sentence if it does not convey any important information.
        \item \textcolor{red}{Merge and shorten}: Merge consecutive sentences if they convey continued information and make the merged sentences concise.
        \item \textcolor{blue}{Rephrase and shorten}: Rephrase a sentence to make it simpler and make it shorter if possible.
    \end{enumerate}    
    \item If the sentences do not express any proper context, or are semantically incorrect, add a comment and proceed to the next sentence.

\end{enumerate}

\textbf{Rules and Tips: }
\begin{itemize}
\sepval
    \item Select all items in the article that have important information.
    \item Make the sentences simpler and concise keeping the important information.
    \item Under each article is a comments box. Optionally, you can provide article-specific feedback in this box. This may include a rationalization of your choice, a description of an error within the article, or the justification of another answer which was also plausible. In general, any relevant feedback would be useful and will help in improving this task.
\end{itemize}

\subsection{Annotator Agreement}
In total nine annotators have annotated the summarization dataset. Each of them is given 70 non-overlapping distinct articles. So there is no disagreement between annotators. Another round of manual cleaning has been done by two meta annotators who have gone through the whole dataset to ensure summarization quality and consistency, by addressing the comments and suggestions made by the annotators in the first round and making necessary changes(update/delete). For example in first round of annotation, annotators put comments like - “The paragraph is vague”, “Independent sentences”, “The paragraph does not have a logical flow. It cannot be further summarized”, “It is not clear what the paragraph is talking about”, etc. These comments are addressed by the meta annotators by manually correcting or removing the articles.

For the classification task, 3 annotators (we call them A1, A2, A3 here) separately annotate the dataset- A1 and A2 annotate 800 examples each and A3 annotates 801 examples. In the second step, they are assigned to reevaluate the annotations of each other (A1 reevaluating labels assigned by A3, A2 reevaluating labels assigned by A1, and A3 reevaluating A2). Such reevaluations bring forth disagreements on several labels which are finally resolved by their combined discussion. For example: “The AMF shall not indicate to the SMF to release the emergency PDU session.”: A2 labels this as Security, while A3 assigns Undefined. This disagreement is later resolved by discussing their reasoning for the respective labels.

\subsection{Examples}
We are listing some example annotated data for both tasks.

\subsubsection{Sentence Classification:} Here we show a few examples of sentence classification--each containing a sentence and the correct label associated with it.\\

\cfexampleheader{1} If the positioning method parameter indicates both E-Cell ID and GNSS positioning, the eNB may \textcolor{red}{use E-Cell ID measurement collection} only if the \textcolor{red}{UE does not provide GNSS-based location information}.
\par ~ \par
\cfexamplelabel{1}\classnameb
\par ~ \par
\cfexampleheader{2} SIGN\_VAR shall be included in the \textcolor{teal}{channel quality report}.
\par ~ \par
\cfexamplelabel{2}\classnamea
\par ~ \par
\cfexampleheader{3} After performing the attach, the MS should \textcolor{red}{activate PDP context(s)} to \textcolor{red}{replace any previously active} PDP context(s).
\par ~ \par
\cfexamplelabel{3} \classnameb
\par ~ \par
\cfexampleheader{4} \textcolor{blue}{It switches} the user from the UTRAN user plane to the GAN user plane
\par ~ \par
\cfexamplelabel{4} \classnamec
\par ~ \par
\cfexampleheader{5} \textcolor{teal}{If the BSIC cannot be decoded at the next available opportunities re attempts shall be made to decode this BSIC.}
\par ~ \par
\cfexamplelabel{5} \classnamea
\par ~ \par
\cfexampleheader{6} \textcolor{blue}{This might lead} to an empty or even absent structure, if \textcolor{blue}{no parameter was modified.}
\par ~ \par 
\cfexamplelabel{6} \classnamec
\par ~ \par

\subsubsection{Summarization:} Here are a few examples, each containing an article and its summary.
\label{apndx:subsec:sumExample}
\par ~ \par
\noindent \textbf{Article 1: } As indicated, 5G NR Meas Gap Length is not fixed and 3GPP specifications made it configurable. Having a fixed Meas Gap could cause unnecessary degradation of throughput in the serving cell. The SMTC window and window duration can be set to match SSB transmissions and accordingly, the MGL. For example, if we consider the SMTC window duration as 2 ms and the Meas Gap Length as 6 ms, here 4 ms segment would not be available for transmission, and reception of data in the serving cell will result in low DL/UL throughput.
\par ~ \par
\noindent \textbf{Summary 1: }5G NR Meas Gap Length is adjustable per 3GPP specs. A fixed Meas Gap can degrade serving cell throughput. The SMTC window and duration can match SSB transmissions and the MGL. If the SMTC window duration is 2 ms and Meas Gap Length is 6 ms, a 4 ms segment is not accessible for transmission, resulting in limited DL/UL throughput.
\par ~ \par
\noindent \textbf{Article 2: } IMSI-catching attacks have threatened all generations (2G/3G/4G) of mobile telecommunication for decades. As a result of facilitating backward compatibility for legacy reasons, this privacy problem appears to have persisted. However, the 3GPP has now decided to address this issue, albeit at the cost of backward compatibility. In case of identification failure via a 5G-GUTI, unlike earlier generations, 5G security specifications do not allow plain-text transmissions of the SUPI over the radio interface. Instead, an Elliptic Curve Integrated Encryption Scheme (ECIES)-based privacy-preserving identifier containing the concealed SUPI is transmitted. This concealed SUPI is known as SUCI (Subscription Concealed Identifier).
\par ~ \par
\noindent \textbf{Summary 2: }Unlike earlier generations, in the case of identification failure via a 5G-GUTI, 5G security specifications do not allow plain-text transmissions of SUPI over the radio interface. Instead, an Elliptic Curve Integrated Encryption Scheme (ECIES)-based privacy-preserving identifier containing the concealed SUPI (also known as SUCI) is transmitted.
\par ~ \par
\noindent \textbf{Article 3: }A SUPI is usually a string of 15 decimal digits. The first three digits represent the Mobile Country Code (MCC) while the next two or three form the Mobile Network Code (MNC), identifying the network operator. The remaining (nine or ten) digits are known as Mobile Subscriber Identification Numbers (MSIN) and represent the individual user of that particular operator. SUPI is equivalent to IMSI, which uniquely identifies the ME, and is also a string of 15 digits.
\par ~ \par
\noindent \textbf{Summary 3: }SUPI is a string of 15 decimal digits consisting of the Mobile Country Code, Mobile Network Code, and Mobile Subscriber Identification Number. SUPI is equivalent to IMSI which uniquely identifies the ME.
\par ~ \par
\noindent \textbf{Article 4: }Next-generation 5G cellular systems will operate in frequencies ranging from around 500 MHz up to 100 GHz. Till now, with LTE and Wi-Fi technologies, we were operating below 6GHz and the channel models were designed and evaluated for operation at frequencies only as high as 6 GHz. The new 5G systems are to operate in bands above 6 GHz and existing channel models will not be valid, hence there is a need for accurate radio propagation models for these higher frequencies, which requires new channel models. The requirements of the new channel model that can support 5G operation across frequency bands up to 100 GHz are based on the existing 3GPP channel models along with extensions to cover additional 5G modeling requirements.
\par ~ \par
\noindent \textbf{Summary 4: }5G will operate in frequencies ranging from around 500 MHz up to 100 GHz. Up to now 4G and WiFi were operating below 6GHz and the channel models were designed and evaluated for operation at frequencies only as high as 6GHz.
\par ~ \par
\noindent \textbf{Article 5: }Carrier Aggregation (CA) increases the bandwidth by combining several carriers. Each aggregated carrier is referred to as a Component Carrier (CC). 5G NR CA supports up to 16 contiguous and non-contiguous CCs with different numerologies in the FR1 band and in the FR2 band. A Carrier aggregation configuration includes the type of carrier aggregation (intra-band, contiguous or not, or inter-band), the number of bands, and the bandwidth class. CA Bandwidth Class is a series of alphabets that defines the minimum and maximum bandwidth along with the number of component carriers. 
\par ~ \par
\noindent \textbf{Summary 5: }Carrier Aggregation (CA) increases the bandwidth by combining several carriers. 5G NR CA supports up to 16 contiguous and non-contiguous CCs with different numerologies.

\subsection{Data Sources}\label{subsec:datasource}
Below we list the websites that were scrapped to create SPEC5G.
\begin{center}

\def\arraystretch{1}
\begin{table*}[ht]
	\centering
	\renewcommand{\arraystretch}{1}
	\fontsize{10}{10}\selectfont
\begin{tabular}{p{0.15\linewidth}p{0.25\linewidth}p{0.3\linewidth}p{0.1\linewidth}p{0.1\linewidth}} 
 \hline 
 \rule{0pt}{2ex}Portal Name & Web Address & Description &\# of Sentences&\# of Words\\ [0.5ex] 
 \hline\hline
 \rule{0pt}{2ex}Artiza Networks & \url{artizanetworks.com}{} & ArtizaNetworks contains tutorials about 3G, 4G, and 5G Radio Access Network (RAN) and Core Network (CN). & 383 & 5182\\
 \hline
 \rule{0pt}{2ex}Event Helix & \url{eventhelix.com}{} & Event Helix is a private corporation based on Maryland. They develop tools for networking and distributed systems and host numerous blogs about 5G Radio, TCP/IP, and so on. & 595 & 6691\\ 
 \hline
 \rule{0pt}{2ex}3G LTE Info & \url{3glteinfo.com}{} & 3G LTE Info offers tutorials and articles for network professionals. These articles encompass GSM, GPRS, 3G, LTE, 5G, Bluetooth, and so on. & 790 & 9651\\
 \hline
 \rule{0pt}{2ex}4G 5G World & \url{4g5gworld.com}{} &  Powered by NgnGuru Solutions Pvt. Ltd., 4G 5G World delivers news, reports, and tutorials about 4G and 5G advanced technologies. & 80 & 1069\\
 \hline
 \rule{0pt}{2ex}Info NR LTE & \url{info-nrlte.com}{} & Run by telecom experts, Info NR LTE delivers technology overviews about NR LTE and NR 5G. & 508 & 7431\\ 
 \hline
 \rule{0pt}{2ex}Resurchify & \url{resurchify.com}{} & Resurchify contains research gatherings from conferences, journals, symposiums, meetings from multiple sectors. & 138 & 1815\\ 
 \hline
 \rule{0pt}{2ex}Share Tech Note & \url{sharetechnote.com}{} & ShareTechNote aims to be a reference guideline on numerous fields, such as, programming languages, engineering, mathematics, advanced technologies. 5G is one of them.  & 8325 & 91575\\ 
 \hline
 \rule{0pt}{2ex}Telecompedia & \url{telecompedia.net}{} &  Telecompedia is a tutorial resource written by 4G, 5G, and radio experts from Rakuten Mobile on different 5G related technologies such as D-RAN, Open-RAN, power control etc. & 2173 & 24997\\ 
 \hline
 \rule{0pt}{2ex}RF Wireless & \url{rfwireless-world.com}{} &  Following IEEE and 3GPP standards, RF Wireless hosts articles, tutorials, source code, terminologies about wireless technologies. & 610 & 8032\\ 
 \hline
 \rule{0pt}{2ex}Tech Play On & \url{techplayon.com}{} & Tech Play On contains technology news and guidelines on 5GNR and LTE. & 5220 & 89035\\ 
 \hline
 \rule{0pt}{2ex}Telecom Hall & \url{telecomhall.net}{} &  A forum to discuss the advances on telecom domain and to guide developers or practitioners. & 7982 & 127712\\ 
 \hline
 \rule{0pt}{2ex}How LTE Stuff Works & \url{howltestuffworks.blogspot.com}{} & Hosts numerous blogs about 5GNR and LTE.  &4213 & 65520 \\
 \hline
 \rule{0pt}{2ex}Pro-Developer Tutorial & \url{prodevelopertutorial.com}{} & Delivers tutorials about C/C++, Git, System design, 4G LTE, 5GNR, shell-scripting, etc. & 3178 & 31708\\ 
 \hline
\end{tabular}
 \caption{List of blogs \& forums crawled as part of dataset collection}
\label{Tab:weblist}
\end{table*}
\end{center}